\documentclass[twocolumn]{autart}    % Enable this line and disable the 
\usepackage{amsmath}
\usepackage{subcaption}
\usepackage{wasysym}
\usepackage{amssymb}
\usepackage{booktabs}
\usepackage{mathtools}
\usepackage{graphicx}          % Include this line if your 
\usepackage{cite}

\newtheorem{thm}{Theorem}
\newtheorem{lem}{Lemma}
\newtheorem{cor}{Corollary}
\newtheorem{assum}{Assumption}

\begin{document}

\begin{frontmatter}

\title{Exact Continuous Reformulations of Logic Constraints in Nonlinear Optimization and Optimal Control Problems} % Title, preferably not more 

%\thanks[footnoteinfo]{The content of this paper was presented at the 64\textsuperscript{th} IEEE Conference on Decision and Control 2025. Corresponding author J.~Wehbeh.}

\author[ICL]{Jad Wehbeh}\ead{j.wehbeh22@imperial.ac.uk},    % Add the 
\author[ICL]{Eric C. Kerrigan}\ead{e.kerrigan@imperial.ac.uk}               % e-mail address 

\address[ICL]{Imperial College London, United Kingdom}

                             % keyword list or with the 
                                          % help of the Automatica 
                                          % keyword wizard

\begin{abstract}                          % Abstract of not more than 200 words.
Many nonlinear optimal control and optimization problems involve constraints that combine continuous dynamics with discrete logic conditions. Standard approaches typically rely on mixed-integer programming, which introduces scalability challenges and requires specialized solvers. This paper presents an exact reformulation of broad classes of logical constraints as binary-variable-free expressions whose differentiability properties coincide with those of the underlying predicates, enabling their direct integration into nonlinear programming models. Our approach rewrites arbitrary logical propositions into conjunctive normal form, converts them into equivalent max–min constraints, and applies a smoothing procedure that preserves the exact feasible set. The method is evaluated on two benchmark problems, a quadrotor trajectory optimization with obstacle avoidance and a hybrid two-tank system with temporal logic constraints, and is shown to obtain optimal solutions more consistently and efficiently than existing binary variable elimination techniques.
\end{abstract}

\end{frontmatter}

\section{Introduction}

In many optimal control problems, system behavior is governed not only by continuous dynamics but also by discrete logic conditions that influence state evolution and constraint activation. Such conditions are naturally encoded using binary or integer variables, and can be solved using a variety of mixed-integer approaches~\cite{wensing2023optimization,malyuta2023fast}. However, mixed-integer optimization can be undesirable or impractical in optimal control contexts, particularly when the system is nonlinear or non-convex, or when the number of discrete variables is small relative to the continuous dynamics. These problems require specialized solvers that may be inaccessible to different applications and introduce significant computational challenges \cite{kronqvist2019review,liberti2019undecidability}. Non-convex continuous dynamics also make it prohibitively expensive for mixed-integer approaches to compute global solutions, effectively removing a key advantage of mixed-integer solution techniques \cite{burer2012non}.
For these reasons, the use of standard continuous nonlinear optimization solvers is generally preferable, as they are widely available and tend to scale more effectively to large problem dimensions. This is often evident with differentiable reformulations of mixed-integer methods, which tend to perform significantly better than the original approach when they exist \cite{kirches2020approximation}. In this paper, we generalize the work of \cite{wehbeh2025smooth} to provide an exact reformulation for general logic constraints as continuous functions with identical solution sets.
 
\subsection{State of the Art}
 
Several approaches have been proposed for modeling logic constraints within the context of optimal control. The simplest formulations describe the logical relations in the problem directly using integer variables that take the value of $1$ when a condition holds and $0$ when it does not. This is usually done by leveraging the big M approach~\cite{bertsimas2021unified},~\cite[Chap.~4]{bazaraa2011linear}  or by introducing appropriate complementarity constraints \cite{belotti2016handling}. However, since these methods introduce discrete optimization variables, their direct solution necessitates the use of mixed-integer solvers. 

Binary variables can be embedded into continuous optimization solvers using equality constraints such as $x(x - 1) = 0$, which enforce $x \in \{0,1\}$, but such methods are known to perform poorly in practice. Alternatively, binary behavior can be approximated using log-sum-exp functions~\cite[Chap.~3]{boyd2004convex}, sigmoid functions~\cite{iliev2017approximation}, or similar techniques. These approximations trade off numerical performance for accuracy and do not guarantee exact logical behavior. In some cases, continuous relaxations of mixed-integer problems can yield feasible solutions without enforcing integrality~\cite{gunluk2010perspective}, but these approaches are limited in scope and are not suitable for the general nonlinear optimal control problems considered here.

Temporal logic constraints offer an alternative for expressing logic in optimization, providing natural representations of time-dependent behavior~\cite{wolff2014optimization}. However, standard methods typically convert such constraints into mixed-integer programs~\cite{kurtz2022mixed}, which suffer from the challenges noted earlier, or employ SMT solvers that perform poorly on nonlinear optimization~\cite{belta2019formal}. Directly applying temporal logic to nonlinear dynamical systems yields intractable non-convex mixed-integer programs, necessitating approximations to the problem~\cite{papusha2016automata}. Constraint programming is another option, but it scales poorly with large numbers of continuous variables~\cite{van1998constraint}.

The most relevant methods to our problem are presented in the recent works of \cite{malyuta2023fast,cafieri2023continuous}, which propose different approaches for handling logical implications of the form $g_1(x) \leq 0 \implies g_2(x) \leq 0$ in the constraints. However, these methods cannot easily deal with more general forms of logic efficiently, as implication lacks functional completeness and forces indirect, expanded representations of basic Boolean operators. In \cite{malyuta2023fast}, the authors integrate implication constraints into a sequential quadratic programming framework using a homotopy map, thus avoiding the need for binary variables. On the other hand, \cite{cafieri2023continuous} introduces a smooth piecewise-quadratic penalty function to enforce the implication constraints directly using continuous optimization solvers and demonstrates the effectiveness of the approach on a feasibility problem.
 
\subsection{Contributions}
 
The main contributions of this paper are as follows:
\begin{itemize}
    \item We introduce a novel method for eliminating binary variables in logical conditions that retains the exactness of the feasible set. If the logical propositions are $r$-differentiable (contained in $\mathcal{C}^r$), then the reformulated logic is also $r$-differentiable.
    \item We provide compact representations of temporal-logic operators such as until and release, whose reformulations introduce only linearly many constraints in the input dimension.
    \item We evaluate our approach on quadrotor trajectory planning with conditional obstacle avoidance, and on hybrid two-tank control with temporal conditions. In both cases, it yields higher success rates and faster solve times than existing differentiable logic formulations.
\end{itemize}
Section~\ref{sec:statement} formalizes the logic-constrained discrete optimization problem. Section~\ref{sec:smooth_reform} then introduces the proposed binary-elimination procedure, and Section~\ref{sec:logic_operators} details its application to common logical operators. Finally, numerical results are presented in Section~\ref{sec:examples}.

\section{Problem Statement}
\label{sec:statement}

Consider a logic-constrained optimization problem with decision variables $z \in \mathbb{R}^n$. The problem can be written
\begin{subequations}
\label{eq:logic_constrained_ocp}
\begin{align}
\label{eq:logic_constrained_ocp_cost}
    \min_{z} \; J(z)& \\ 
\text{s.t.}  \qquad g(z) &\leq 0 \\
\label{eq:logic_constrained_ocp_eqs}
h(z) &= 0 \\
\label{eq:logic_constr_true}
L(z) \ &\text{is}\ \textbf{true}
\end{align}
\end{subequations}
where $J(\cdot) : \mathbb{R}^n \rightarrow \mathbb{R}$ is the cost function to be minimized, $g(\cdot) : \mathbb{R}^n \rightarrow \mathbb{R}^{n_g}$ is a set of inequality constraints that the solution to the optimization problem must satisfy, $h(\cdot) : \mathbb{R}^n \rightarrow \mathbb{R}^{n_h}$ is a similar set of equality constraints, and $L(\cdot) : \mathbb{R}^n \rightarrow \{\textbf{true},\textbf{false}\}$ is a logic program whose output is required to be \textbf{true}. The constraint of~\eqref{eq:logic_constr_true} enables the encoding of any non-differentiable requirements on the evolution of $z$, and augments the standard smooth optimization formulations of \eqref{eq:logic_constrained_ocp_cost} through \eqref{eq:logic_constrained_ocp_eqs}. The resulting description is general enough to capture optimal control problems by treating the state $x$ and the control inputs $u$ as part of $z = (x,u)$, as illustrated in the examples of Section \ref{sec:examples}. When doing so, the system dynamics are added to $h(\cdot)$, or to $L(\cdot)$ if their evolution relies on logic. 

We now formalize the allowable forms of $L(\cdot)$. 
Define an equality proposition $Q(\cdot) : \mathbb{R}^n \rightarrow \{\textbf{true},\textbf{false}\}$, which evaluates to \textbf{true} if and only if its associated function $q(\cdot)$ satisfies $q(z) = 0$, where $q(\cdot) : \mathbb{R}^n \rightarrow \mathbb{R}$. 
Similarly, define an inequality proposition $P(\cdot) : \mathbb{R}^n \rightarrow \{\textbf{true},\textbf{false}\}$, which evaluates to \textbf{true} whenever its associated function $p(\cdot) : \mathbb{R}^n \rightarrow \mathbb{R}$ satisfies $p(z) \leq 0$, and \textbf{false} otherwise. 

\begin{assum}
\label{assum:prob_form}
Let $\mathcal{L}$ denote the set of all logic problems consisting of $n_q \in \mathbb{N}_0$ equality propositions $Q_i$, $i \in \{1,\ldots,n_q\}$, and $n_p \in \mathbb{N}_0$ inequality propositions $P_j$, $j \in \{1,\ldots,n_p\}$. 
These propositions are combined using finitely many logical operators—NOT ($\neg$), AND ($\land$), and OR ($\lor$)—to output a single Boolean value. For the remainder of this paper, let $L \in \mathcal{L}$.
\end{assum}

Allowing the operators $\neg$, $\land$, and $\lor$ in the definition of $L$ is sufficient to express any boolean relation among propositions, since this set of operators is functionally complete~\cite[Thm~15D]{enderton2001mathematical}. Together with equality and inequality propositions, this implies that $\mathcal{L}$ can represent arbitrary requirements on $x$ and $u$. In principle, $L$ could also be structured to incorporate the constraints $g$ and~$h$. However, we choose to treat these separately in order to distinguish conventional smooth constraints from those that inherently require a logical structure.

Throughout this work, our goal is to replace the logical constraint in~\eqref{eq:logic_constr_true} with binary-free equality and inequality constraints that can be embedded into $g$ and $h$ without altering the feasible set of Problem~\eqref{eq:logic_constrained_ocp}. This yields a standard continuous optimal control problem that is solvable with conventional nonlinear programming methods.

\section{Continuous Constraint Reformulation}
\label{sec:smooth_reform}

In this section, we demonstrate that, under some mild conditions, any logical problem satisfying Assumption~\ref{assum:prob_form} admits a continuous reformulation with an identical solution set. To that end, we begin by introducing the following result, whose proof illustrates the constructive procedure used in rewriting such problems.
 
\subsection{Logic Simplification}
 
\begin{lem}
\label{lemma:no_neg_eq}
 Any problem in $\mathcal{L}$ can be rewritten using only inequality propositions,
$\land$, and $\lor$ operators.
\end{lem}
\begin{pf}
    We first show that equality propositions can be represented equivalently by inequality propositions. To do so, we introduce an additional pair of inequalities
    \begin{subequations}
    \begin{align}
        P^+_{n_p +i} \land P^-_{n_p +i} \equiv& \; Q_i \qquad \qquad \forall i \in \{1,\ldots,n_q\} \\
        p_{n_p+i}^+(z) \coloneqq& \;q_i(z) \\
        p_{n_p+i}^-(z) \coloneqq& \;-q_i(z)
    \end{align}
    \end{subequations}
    so that $P^+_{n_p+i}$ enforces $q_i(z)\le 0$ and $P^-_{n_p+i}$ enforces $-q_i(z)\le 0$. Together, these hold if and only if $q_i(z) = 0$. 
    Next, we eliminate negations by pushing the $\neg$ operators down to the proposition level using De Morgan’s laws:
    \begin{subequations}
    \begin{align}
        \neg (P_1 \land P_2) & = \neg P_1 \lor \neg P_2 \\   
        \neg (P_1 \lor P_2) &= \neg P_1 \land \neg P_2 \\
        \neg \neg P &= P.
    \end{align}
    \end{subequations}
    This allows any problem in $\mathcal{L}$ to be expressed using $\land$, $\lor$, and $n_p + 2n_q $ (possibly negated) inequality propositions.
    We then fully eliminate the negations by noticing that
    \begin{equation}
    \begin{split}
        &\neg[p_i(z) \leq 0] \iff [p_i(z) > 0] \\ & \qquad \qquad\forall z \in \mathbb{R}^n,\, i \in \{1,\ldots,n_p  + 2n_q \}.
    \end{split}
    \end{equation}
    For any strict inequality, we know that
    \begin{equation}
    \label{eq:strict_to_nonstrict}
        p_i(z) > 0 \; \iff \; 
\exists\,\eta_i \in 
\mathbb{R}:\; -p_i(z) + \exp(\eta_i) \le 0
    \end{equation}
    Define the negated proposition
    \begin{subequations}
    \label{eq:neg_ineq_equiv}
    \begin{align}
    \bar{P}_i & \equiv \neg P_i 
    \qquad \forall i \in \{1,\ldots,n_p  + 2n_q \} \\
    \bar{p}_i(z,\eta_i) &\coloneqq  - p_i(z) + \exp(\eta_i).
\end{align}
\end{subequations}
From \eqref{eq:strict_to_nonstrict}, introducing an auxiliary variable $\eta_i$ for each negated inequality allows satisfying the condition $\neg P_i$ by enforcing $\bar{p}_i(z,\eta_i)\le 0$. To maintain the structure of Assumption \ref{assum:prob_form}, the optimization problem's decision vector can then be augmented by a vector of these new auxiliary variables $\eta$, and redefined as $z' = (z,\eta)$. 

Consequently, any problem in $\mathcal{L}$ can be reformulated to only use inequality propositions, $\land$, and $\lor$ operators \qed
\end{pf}

In finite-precision solvers, the variables $\eta_i$ must be kept within bounds to avoid numerical overflow or underflow. Moreover, constraints are only satisfied up to a prescribed tolerance. Consequently, in practical applications it typically suffices to choose $\bar{p}_i = -p_i$, or to fix $\exp(\eta_i)$ to a value smaller than the optimizer’s tolerance, thereby eliminating the need to explicitly include the auxiliary variable $\eta$. For these reasons, and to keep the notation streamlined, we continue to let $z$ represent the full decision vector throughout the rest of this work. If one needs the feasibility sets to be exactly equivalent under infinite precision, then $z$ can be interpreted as also containing the auxiliary variable vector $\eta$.
 
\subsection{Max-Min Reformulation}
 
Next, we attempt to move away from the logical operators $\land$ and $\lor$, which cannot inherently be represented in continuous optimization solvers, and replace them with a formulation using the continuous (but not necessarily differentiable) $\max$ and $\min$ operators.

\begin{lem}
\label{lemma:constraint_equiv_minmax}
    For any problem $L \in \mathcal{L}$, and for any choice of $z \in \mathbb{R}^n$ the following statements are equivalent:
    \begin{itemize}
        \item The constraint of \eqref{eq:logic_constr_true} holds
        \item
\begin{minipage}{\linewidth}
\begin{equation}
\label{eq:logic_constr_cnf}
    \bigwedge_{i=0}^{n_\land} \bigvee_{j=0}^{n_\lor(i)} P_{i,j} \ \text{is} \ \textbf{true}
\end{equation}
\end{minipage}
        where \eqref{eq:logic_constr_cnf} is the conjunctive normal form (CNF) of $L$ under Lemma \ref{lemma:no_neg_eq}. This form has $n_\land+1$ clauses joined by~$\land$ operators, each containing $n_\lor(i)+1$ inequality propositions connected by $\lor$ operators, where $i \in \{0,\ldots,n_\land\}$ corresponds to the index of the clause.
        \item \begin{minipage}{\linewidth}
\begin{equation}
\label{eq:logic_constr_maxmin}
\max_{i \in \{0,\ldots,n_\land\}} \ \min_{j \in \{0,\ldots,n_{\lor}(i)\}} p_{i,j}(z) \leq 0
\end{equation}
\end{minipage}
        where $p_{i,j}$ in \eqref{eq:logic_constr_maxmin} corresponds to $P_{i,j}$ in \eqref{eq:logic_constr_cnf} $\forall i \in \{0,\ldots,n_\land\}$, $ j \in\{0,\ldots,n_{\lor} (i)\}$.
    \end{itemize}
\end{lem}
\begin{pf}
    The equivalence between \eqref{eq:logic_constr_true} and \eqref{eq:logic_constr_cnf} is relatively straightforward.  By Lemma~\ref{lemma:no_neg_eq}, $L$ can be expressed without $\neg$ operators and without equality propositions. Moreover, any propositional formula admits an equivalent CNF (a conjunction of disjunctions, or AND or ORs) that preserves truth values \cite[Chap.~1]{enderton2001mathematical}. Thus, \eqref{eq:logic_constr_true} can always be rewritten as \eqref{eq:logic_constr_cnf}.  

    We now show that \eqref{eq:logic_constr_cnf} and \eqref{eq:logic_constr_maxmin} are equivalent. Consider a single clause $i$ defined by
\begin{equation}
\left[p_{i,0}(z)\leq 0\right] \lor \cdots \lor \left[p_{i,n_\lor(i)}(z)\leq 0\right].    
\end{equation}
This clause is \textbf{true} if and only if at least one inequality holds, i.e., iff
\begin{equation}
\min_{j \in \{0,\ldots,n_\lor(i)\}} p_{i,j}(z) \leq 0.    
\end{equation}
Analogously, the conjunction of multiple clauses
\begin{equation}
\bigwedge_{i=0}^{n_\land} \left(\min_{j} p_{i,j}(z) \leq 0\right)    
\end{equation}
is true if and only if the largest of these inner minima also satisfies the inequality, i.e.,
\begin{equation}
    \max_{i \in \{0,\ldots,n_\land\}} \ \min_{j \in \{0,\ldots,n_\lor(i)\}} p_{i,j}(z) \leq 0.
\end{equation}
This establishes the equivalence between \eqref{eq:logic_constr_cnf} and \eqref{eq:logic_constr_maxmin}. Combining both steps completes the proof. \qed
\end{pf}
Recall that the total number of distinct propositions appearing in \eqref{eq:logic_constr_cnf} and \eqref{eq:logic_constr_maxmin} is $n_p + 2n_q$. The conversion to conjunctive normal form does not introduce any new propositions, only rearranging or duplicating existing ones. Consequently, the notation $P_{i,j}$ should be interpreted as a simple renumbering of the propositions for notational convenience.
 
\subsection{Constraint Smoothing}
 
In order to enable the smoothing of \eqref{eq:logic_constr_maxmin}, we first introduce the following assumption.

\begin{assum}
\label{assum:smoothing}
The functions $p_{i,j} \in \mathcal{C}^r$ ($r$-times continuously differentiable, $r \in \{0, \ldots, +\infty\}$) for all indices
$i \in \{0, \ldots, n_\land\}$ and $j \in \{0, \ldots, n_\lor(i)\}$.  
\end{assum}

Assumption \ref{assum:smoothing} is not restrictive and simply provides an order of differentiability for the proof of Theorem \ref{thrm:smoothing_result}. For many problems, the order of differentiability $r$ can be increased by introducing appropriate logic to break down the logical propositions at the non-differentiable points. 

\begin{thm}
\label{thrm:smoothing_result}
Let a constraint be of the form \eqref{eq:logic_constr_true}, and suppose that $L \in \mathcal{L}$ satisfies Assumptions \ref{assum:prob_form} and \ref{assum:smoothing}. Then, this constraint can be equivalently rewritten as
\begin{equation}
\label{eq:logic_constr_smooth}
    \sum_{j=0}^{n_\lor(i)} \lambda_{i,j} \, p_{i,j}(z) \leq 0, \quad \forall i \in \{0, \ldots, n_\land\},
\end{equation}
without changing the feasible solution set. Here,
\begin{itemize}
    \item $\lambda_i := (\lambda_{i,0}, \lambda_{i,1}, \dots, \lambda_{i,n_\lor(i)}) \in \Lambda_{n_\lor(i)+1}$ is a collection of auxiliary smoothing variables $\forall i \in \{0, \dots, n_\land\}$, where $\Lambda_k$ denotes the probability simplex in $\mathbb{R}^k$:
    \[
    \Lambda_k \coloneqq \Big\{ l \in \mathbb{R}^k \,\big|\, \sum_{i=1}^k l_i = 1,\, l_i \ge 0 \Big\}.
    \]
    \item $p_{i,j}(z)$, $n_\land$, and $n_\lor(\cdot)$ are as defined in \eqref{eq:logic_constr_maxmin}.
\end{itemize}
Moreover, the constraint in \eqref{eq:logic_constr_smooth} is differentiable up to the order of differentiability of each function $p_{i,j}$, and is therefore at least $r$-differentiable.
\end{thm}
\begin{pf}
    Lemma \ref{lemma:constraint_equiv_minmax} allows us to replace the constraint of \eqref{eq:logic_constr_true} with that of \eqref{eq:logic_constr_maxmin} without modifying the solution set. Then, under Assumption \ref{assum:smoothing}, we can apply the smoothing approach of \cite{kirjner1998conversion} to separate the finite $\max$ into $n_\land +1$ different constraints and smooth each $\min$ to obtain the form of \eqref{eq:logic_constr_smooth}. Following the main result of \cite{kirjner1998conversion}, this set of smoothed constraints has the same feasible region as \eqref{eq:logic_constr_maxmin}, proving the equivalence of~\eqref{eq:logic_constr_true} and \eqref{eq:logic_constr_smooth}. Under Assumption \ref{assum:smoothing}, the mapping $(z,\lambda)\mapsto \sum_{j=0}^{n_\lor(i)} \lambda_{i,j} \, p_{i,j}(z)$ is $\mathcal{C}^r$ for all $i$. \qed
\end{pf}
 
\subsection{Binary-Free Optimal Control Problem Formulation}
 
By using Theorem \ref{thrm:smoothing_result}, problems of the form \eqref{eq:logic_constrained_ocp} can be transformed into equivalent problems
\begin{subequations}
\label{eq:smooth_ocp}
\begin{align}
    \min_{z,\lambda} \; J(z) & \\
\text{s.t.} \qquad g^*(z,\lambda) &\leq 0 \\
h^*(z,\lambda) &= 0
\end{align}
\end{subequations}
where $g^*(\cdot,\cdot) : \mathbb{R}^n \times \prod_{i = 0}^{n_\land} \Lambda_{n_\lor(i)+1} \rightarrow \mathbb{R}^{n_g^*}$ combines the constraints of $g$ with the reformulated logic constraints and the positivity constraints for the $\lambda$ variables, $h^*(\cdot,\cdot) : \mathbb{R}^n \times \prod_{i = 0}^{n_\land} \Lambda_{n_\lor(i)+1} \rightarrow \mathbb{R}^{n_h^*}$ combines the constraints of $h$ with the unity constraints in the definition of the $\Lambda$ sets, $n_g^* \coloneqq n_g + n_\land + 1 +\sum_{i = 0}^{n_\land} (n_\lor(i)+1)$, and $n_h^* \coloneqq n_h + n_\land + 1$.

This problem formulation is $r$-differentiable and can be handled directly by standard continuous optimization algorithms. The resulting constraints $g^*$ are generally non-convex, yielding multiple local minima; however, this non-convexity stems from the logic itself rather than from the reformulation. Importantly, local minima that satisfy the logic can be readily distinguished from infeasible ones, allowing feasible trajectories to be found effectively through multistart methods. We also note that CNF representations are not unique, and some choices may offer substantially better performance in practice, as illustrated in Section~\ref{sec:logic_operators}.

\section{Common Logical Operators}
\label{sec:logic_operators}
In this section, we describe how some common temporal-logic operators can be expressed in CNF, enabling efficient representations of the form of \eqref{eq:logic_constr_cnf}. To that end, we introduce two logical propositions, $A$ and $B$, which can be tested at any instance of the discrete finite horizon problem between time steps $0$ and $N$. We use $A[i]$ to denote the proposition $A$ tested at time step $i$.
 
\subsection{Implication Operator}
 
The implication operator, $A[i] \implies B[j]$, $i,j \in \{0,\ldots,N\}$, requires that $B[j]$ holds if $A[i]$ holds. It is easily used within our framework, since its equivalent form using basic connectives is already in CNF:
\begin{equation}
    \neg A[i] \lor B[j]
\end{equation}
 
\subsection{Equivalence Operator}
 
The equivalence operator, $A[i] \leftrightarrow B[j]$, $i,j \in \{0,\ldots,N\}$, requires that $B[j]$ holds if and only if $A[i]$ also holds. This can be rewritten as $(A[i] \implies B[j]) \land (B[j] \implies A[i])$, which has the equivalent CNF
\begin{equation}
    (\neg A[i] \lor B[j]) \land (\neg B[j] \lor A[i])
\end{equation}

%\subsection{Always Operator}
%The always operator, $(\mathcal{G} A)[i]$, $i \in \{0,\ldots,N\}$, requires that $A[j]$ hold $\forall j \in \{i,\ldots,N\}$. Since the $\forall$ condition occurs over a finite set, it can be replaced by a finite AND and expressed as
%\begin{equation}
%    \bigwedge_{j=i}^N A[j]
%\end{equation}
 
\subsection{Until Operator}
 
\label{sec:until_operator}
The until operator, $(A \ \mathcal{U} \, B)[i]$, $i \in \{0,\ldots,N\}$, is commonly used in linear temporal logic and requires that from the starting time step $i$, $A$ must remain $\mathbf{true}$ until $B$ is $\mathbf{true}$, and that $B$ must take a $\mathbf{true}$ value by time step~$N$. The until operator is formally defined as
    \begin{equation}
        (A \ \mathcal{U} \, B)[i] \;\;\equiv\;\; \exists j \geq i, \, j \leq N \;:\; B[j] \land \bigwedge_{k=i}^{j-1} A[k].
        \label{eq:U_Semantic}
    \end{equation}
This converts to $(N-i + 1)^2$ disjunctive clauses if expressed in CNF naively. Instead, we assert the following.
\begin{lem}
$(A \ \mathcal{U} \, B)[i]$ can be represented by $N-i + 1$ clauses as
    \begin{equation}
        \bigwedge_{j=i}^{N-1}\; \left( A[j] \;\;\lor\;\; \bigvee_{k=i}^{j} B[k] \right) \;\;\land\;\; \bigvee_{j=i}^N B[j].
        \label{eq:U_CNF}
    \end{equation}
\end{lem}
\begin{pf}
    We first demonstrate how \eqref{eq:U_Semantic} implies \eqref{eq:U_CNF}. From \eqref{eq:U_Semantic}, there exists a smallest $j^*$ between $i$ and $N$ such that $B[j^*]$ is $\mathbf{true}$, and $A[j]$ must hold $\forall j < j^*$. The existence of $B[j^*]$ implies that $\bigvee_{j=i}^N B[j]$ holds, and the rest of \eqref{eq:U_CNF} similarly follows since $A[j]$ is $\mathbf{true}$ for $j < j^*$, and $\bigvee_{k=i}^{j} B[k]$ is $\mathbf{true}$ for $j \geq j^*$. 

    Next, we show why \eqref{eq:U_CNF} implies \eqref{eq:U_Semantic}, and the two forms are consequently equivalent. As before, the existence of a smallest $j^*$ such that $B[j^*]$ is $\mathbf{true}$ follows from the term $\bigvee_{j=i}^N B[j]$. We also know that $\forall j < j^*$, the first part of \eqref{eq:U_CNF} implies that $A[j]$ must be $\mathbf{true}$, since $\bigvee_{k=i}^{j} B[k]$ is $\mathbf{false}$ by the definition of $j^*$. Therefore, \eqref{eq:U_Semantic} holds at~$j^*$, satisfying the existence requirement.  
    \qed
\end{pf}
 
\subsection{Release Operator}
 
The release operator, $(A \ \mathcal{R} \ B)[i]$, $i \in \{0,\ldots,N\}$, requires that starting from time step $i$, $B$ remain $\mathbf{true}$ up to and including the point when $A$ becomes $\mathbf{true}$. This is formally expressed as
    \begin{equation}
        (A \ \mathcal{R} \ B)[i] \;\;\equiv\;\; \forall j \geq i, \, j \leq N \;:\; B[j] \lor \bigvee_{k=i}^{j-1} A[k].
        \label{eq:R_Semantic}
    \end{equation}
By converting the $\forall$ operator into a logical AND, we obtain the CNF
    \begin{equation}
        \bigwedge_{j=i}^N \left( B[j] \lor \bigvee_{k=i}^{j-1} A[k] \right).
    \end{equation}

\section{Numerical Examples}
\label{sec:examples}

\begin{figure*}[bt]
  \centering
  \begin{tabular}{ c @{\hspace{10pt}} c @{\hspace{10pt}} c}
    \includegraphics[height=128pt,trim = {0 10pt 0 20pt}]{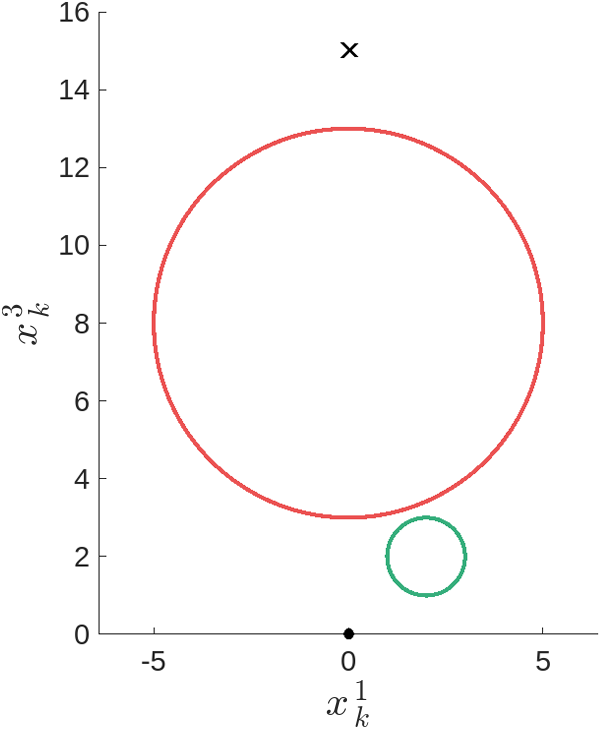} \hspace{20pt} &
    \includegraphics[height=128pt,trim = {0 10pt 0 20pt}]{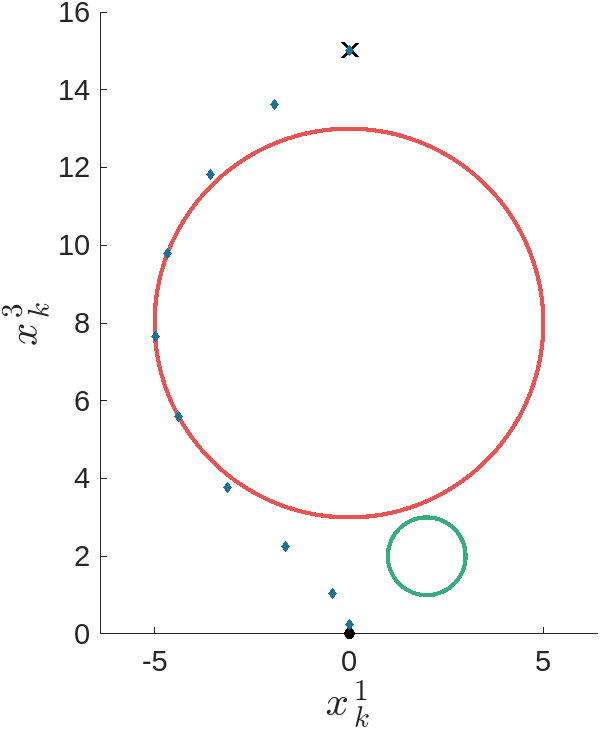} \hspace{20pt} & 
    \includegraphics[height=128pt,trim = {0 10pt 0 20pt}]{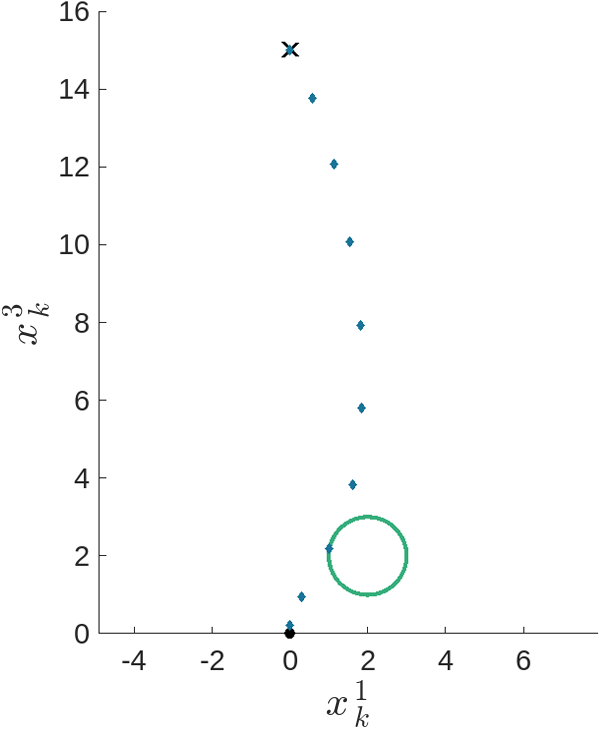} \\[5pt]
    \parbox[b]{.64 \columnwidth}{\footnotesize (a) Going through the green circle allows passing through the red obstacle.} & \hspace{10pt}
      \parbox[b]{.62 \columnwidth}{\footnotesize (b) Sample sub-optimal trajectory. \phantom{filler} \phantom{filler} \phantom{filler} \phantom{filler}} &
      \hspace{10pt} \parbox[b]{.54 \columnwidth}{\footnotesize (c) Optimal solution trajectory. \phantom{filler} \phantom{filler} \phantom{filler} \phantom{filler}}
  \end{tabular}
   
  \caption{Obstacle (red), logic trigger (green), target (black), and solution trajectory (blue) for Problem 1.}
  \label{fig:prob_1}
\end{figure*}
 
\subsection{Quadrotor Control}
 
\begin{figure}[t]
    \centering
    \includegraphics[width=0.45\columnwidth]{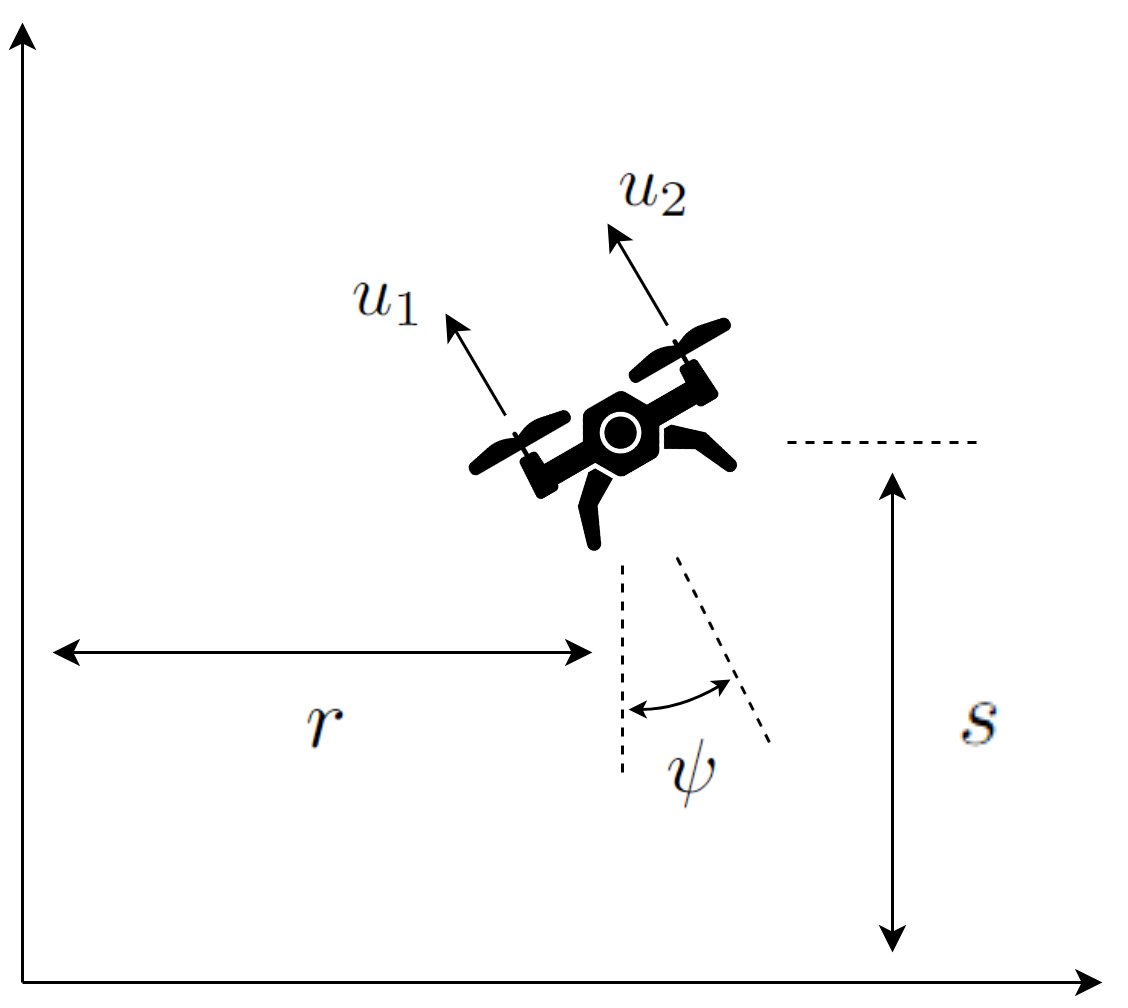}
     
    \caption{Illustration of the quadrotor's horizontal position ($r$), altitude ($s$), and tilt angle ($\psi$) taken from \cite{wehbeh2024robust}.}
    \label{fig:quad_example}
\end{figure}

Consider the quadrotor control problem with dynamics
\begin{equation}
\label{eq:ex_state_dyn}
    \begin{bmatrix}
        {x}^1_{k+1} \\
        {x}^2_{k+1} \\
        {x}^3_{k+1} \\
        {x}^4_{k+1} \\
        {x}^5_{k+1} \\
        {x}^6_{k+1} 
    \end{bmatrix}
    =
    \begin{bmatrix}
        {x}^1_{k} \\
        {x}^2_{k} \\
        {x}^3_{k} \\
        {x}^4_{k} \\
        {x}^5_{k} \\
        {x}^6_{k} 
    \end{bmatrix} +
    T_s
    \begin{bmatrix}
    \left(x^2_{k} + x^2_{k+1} \right)/2 \\[3pt]
    \sin(x_k^5) \left(u^1_{k+1} + u_{k+1}^2 \right) /\gamma \\[3pt]
    \left(x^4_{k} + x^4_{k+1} \right)/2 \\[3pt]
    \cos(x_k^5) \frac{\left(u^1_{k+1} + u_{k+1}^2 \right)}{\gamma} - g_{\earth} \\[3pt]
    \left(x^6_{k} + x^6_{k+1} \right)/2 \\[3pt]
    \ell (u^1_{k+1} -u^2_{k+1}) / I
    \end{bmatrix}
\end{equation}
where $x^1_k$ through $x^6_k$ are the discrete-time quadrotor states corresponding to the continuous values of $r$, $\dot{r}$, $s$, $\dot{s}$, $\psi$, and $\dot{\psi}$ respectively, as shown in Figure \ref{fig:quad_example}. Here, $u_k^1$ and $u_k^2$ are the thrusts at time step $k$, $T_s = 0.25\,\text{s}$ is the discretization time step, $\gamma = 0.15\,\text{kg}$ is the vehicle's mass, $I = 0.00125\,\text{kg.m}^2$ is the moment of inertia, $\ell = 0.1\,\text{m}$ is the motor moment arm, and $g_{\earth} = 9.81\,\text{m/s}^2$ is the gravity acting on the system. The thrusts produced are constrained such that $0 \leq u_k^i \leq 2 \; \forall i \in \{1,2\}, \, k \in \{1,\ldots,N\}$
over the $N = 10$ predicted time steps. The system is initialized from $x_0 = [0,0,0,0,0,0]$, and we attempt to minimize the objective
\begin{equation}
    \sum_{k=1}^N \; (u_k^1)^2 + (u_k^2)^2. 
\end{equation}
The problem is subject to the terminal constraints $x_N^1 = 0$ and $x_N^3 = 15$, and must satisfy the logic program
\begin{equation}
\begin{split}
    & \neg \bigvee_{i = \{2,3\}} \left\{ (x_i^1 - 2)^2 + (x_i^3 - 1)^2 - 1 \leq 0   \right\} \\ & \implies  \neg \bigvee_{i = \{5,\ldots,9\}} \left\{ (x_i^1)^2 + (x_i^3 - 8)^2 - 25 \leq 0 \right\} 
\end{split}
%\raisetag{25pt}
\end{equation}
which, as illustrated in Figure \ref{fig:prob_1}, forces the quadrotor to avoid the red circle on its way to its target unless it passes through the green circle at time steps 2 or 3. Using the procedure of Sections \ref{sec:smooth_reform} and \ref{sec:logic_operators}, this constraint can be smoothed into
\begin{multline}
    \lambda_{1} \left( (x_2^1 - 2)^2 + (x_2^3 - 1)^2 - 1  \right) \\
    + \lambda_{2} \left( (x_3^1 - 2)^2 + (x_3^3 - 1)^2 - 1  \right)  \\
    - \lambda_{3} \left( (x_i^1)^2 + (x_i^3 - 8)^2 - 25  \right ) \leq 0, 
\end{multline}
for all $i \in \{5,\ldots,9\}$, where $\lambda \in \Lambda_3$. The same smoothing variables $\lambda$ can be used $\forall i$, because if the green circle condition is met for one instance, then none of the red circle conditions need apply. The problem is then solved in Julia using the JuMP \cite{Lubin2023} package and the Ipopt \cite{wachter2006implementation} optimizer on a laptop with an 11th Gen Intel\textsuperscript{\tiny\textregistered} Core\textsuperscript{\texttrademark} i7-11370H CPU at 3.30 GHz and 16GB of RAM. We compare our solution method to an approach that enforces the binary variables through a complementarity condition, and to a Big M method that represents the condition $\{a \leq 0\} \lor \{b \leq 0\} \lor \{c \leq 0\}$ as $\{a \leq \mu_1M\} \land \{b \leq \mu_2 M\} \land \{c \leq \mu_3 M\}$ with continuous variables $\mu$ such that $\mu_1 \mu_2 \mu_3 = 0$.

\begin{table}[tb]
\centering
\caption{Performance Metrics on Problem 1 }
\label{tab:prob_1_perf}
\begin{tabular}{lcccc}
\hline
                & Opt. \# & Sub-Opt.  \# & Inf. \# & Avg. Cost \\ \hline
Smoothed        & 813         & 143              & 44                 & 22.24     \\
Compl. & 325         & 518              & 157                & 28.67     \\
Big M           & 660         & 274              & 66                 & 23.73     \\ \hline
                & Avg. Time & \multicolumn{2}{c}{Avg. Time (Feas.)} & Max Time \\ \hline
Smoothed        & 22.4\,ms     & \multicolumn{2}{c}{19.4\,ms}              & 466.8\,ms   \\
Compl. & 76.1\,ms      & \multicolumn{2}{c}{78.2\,ms}              & 419.7\,ms  \\
Big M           & 56.7\,ms      & \multicolumn{2}{c}{37.9\,ms}              & 1782.4\,ms   \\ \hline
\end{tabular}
\end{table}

Since the problem is non-convex in all cases, and our solution method is tested using a local optimizer, we evaluate the solution approaches using 1000 initializations randomly selected from a uniform distribution over the optimization variable limits. The results, detailed in Table \ref{tab:prob_1_perf}, show that our method outperforms the other approaches, returning the optimal solution (shown in Figure \ref{fig:prob_1} (c)) 81.3\% of the time and failing to produce a feasible trajectory in only 4.4\% of runs. The smoothed constraints also display the fastest computation times, even when the slower infeasible runs are excluded. Once again, we remark that sub-optimal outputs are observed despite the exactness of the smoothing due to the non-convex nature of the optimization problem. Using a global optimization solver with our proposed constraint formulation would guarantee optimal trajectories.

\begin{figure}[t]
    \centering
    \includegraphics[width=0.55\columnwidth]{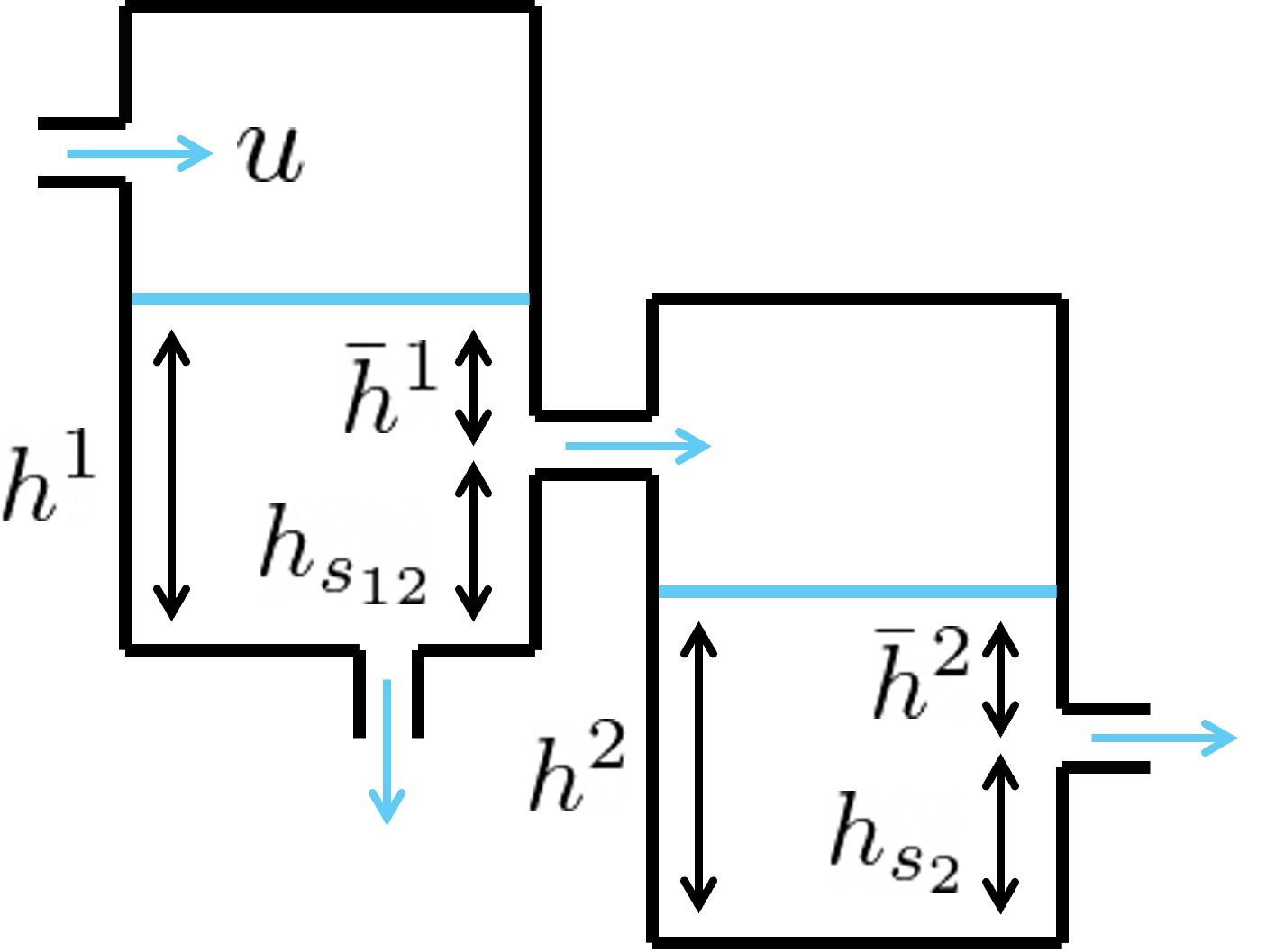}
    \caption{Two tank problem setup showing measurements (in black) and water flow (in blue).}
    \label{fig:two_tank_diagram}
\end{figure}
 
\subsection{Two-Tank Problem}
 
\begin{figure*}[t]
    \centering
    % First subfigure
    \begin{subfigure}{0.32\textwidth}
        \includegraphics[width=\linewidth]{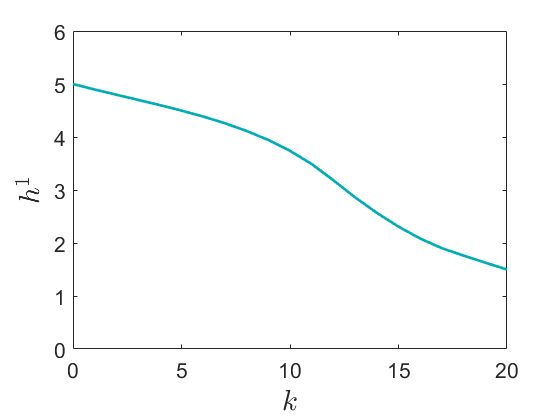}
        %\caption{First}
    \end{subfigure}
    \hfill
    % Second subfigure
    \begin{subfigure}{0.32\textwidth}
        \includegraphics[width=\linewidth]{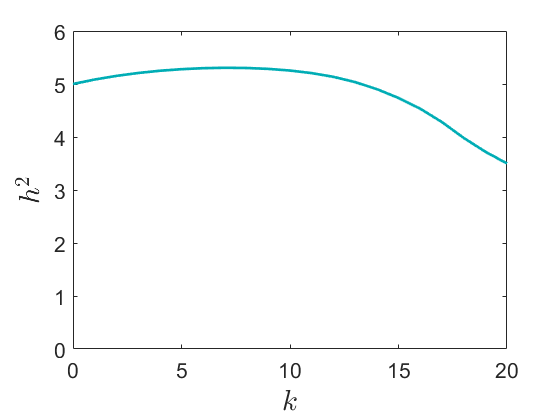}
        %\caption{Second}
    \end{subfigure}
    \hfill
    % Third subfigure
    \begin{subfigure}{0.32\textwidth}
        \includegraphics[width=\linewidth]{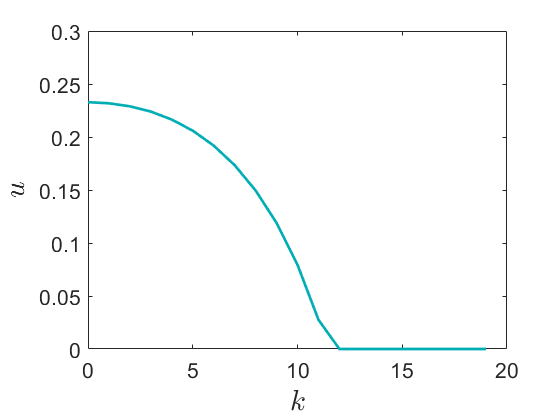}
        %\caption{Third}
    \end{subfigure}
    \vspace{-8pt}
    \caption{Optimal Trajectory for Case 1.}
    \label{fig:two_tank_1}
\end{figure*}
\begin{figure*}[t]
    \centering
    % First subfigure
    \begin{subfigure}{0.32\textwidth}
        \includegraphics[width=\linewidth]{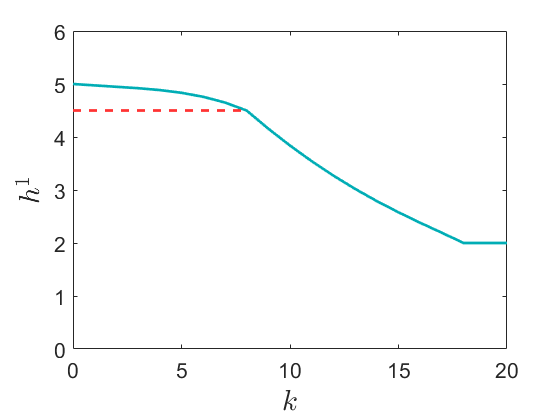}
        %\caption{First}
    \end{subfigure}
    \hfill
    % Second subfigure
    \begin{subfigure}{0.32\textwidth}
        \includegraphics[width=\linewidth]{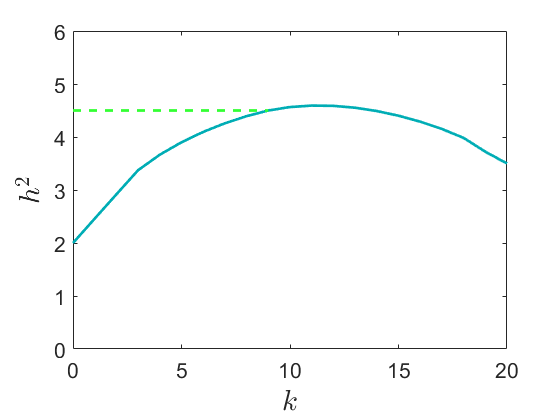}
        %\caption{Second}
    \end{subfigure}
    \hfill
    % Third subfigure
    \begin{subfigure}{0.32\textwidth}
        \includegraphics[width=\linewidth]{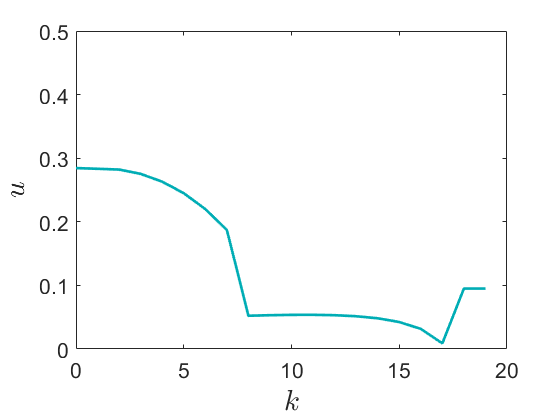}
        %\caption{Third}
    \end{subfigure}
    \vspace{-8pt}
    \caption{Optimal Trajectory for Case 2 with the condition on $h^2$ (green) that releases the constraint on $h^1$ (red).}
    \label{fig:two_tank_2}
\end{figure*}

Next, we consider a version of the two-tank problem shown in Figure \ref{fig:two_tank_diagram} with discrete-time dynamics
\begin{subequations}
\begin{align}
    \begin{split}
        h^1_{k+1} &= h^1_{k} \,+ \\&\;\; T_s\left(\frac{u_{k+1}}{A_1} - \frac{s_{1}\sqrt{2g_{\earth}h^{1\vphantom{+}}_{k}}}{A_1} -\frac{s_{12}\sqrt{2g_{\earth}\bar{h}^1_{k}}}{A_1}\right)
    \end{split} \\
    h^2_{k+1} &= h^2_{k} + T_s\left( \frac{s_{12}\sqrt{2g_{\earth}\bar{h}^1_{k}}}{A_2} -\frac{s_{2}\sqrt{2g_{\earth}\bar{h}^{2}_{k}}}{A_2}\right)
\end{align}    
where $h^1_k$ and $h^2_k$ are the levels of water in tanks $1$ and~$2$ at time step $k$, $u_k$ is the water flow into tank 1, and $A_1$ and $A_2$ are the cross-sectional areas of the tanks. We use $s_1$ to refer to the area of the opening at the bottom of tank $1$, $s_{12}$ for that between tanks $1$ and $2$, and $s_{2}$ for the outlet to tank $2$. We also define
\begin{align}
\label{eq:h1_bar}
\bar{h}^1_k &=
\begin{cases}
h^1_k - h_{s_{12}} & \text{if } h^1_k \geq h_{s_{12}}, \\[2pt]
0 & \text{otherwise},
\end{cases} \\[6pt]
\label{eq:h2_bar}
\bar{h}^2_k &=
\begin{cases}
h^2_k - h_{s_{2}} & \text{if } h^2_k \geq h_{s_{2}}, \\[2pt]
0 & \text{otherwise}.
\end{cases}
\end{align}
to be the heights of the water levels above the respective outlets, where $h_{s_{12}}$ and $h_{s_{2}}$ are defined as in Figure \ref{fig:two_tank_diagram}. The parameter values used are specified in Table \ref{tab:two_tank_param}.
\end{subequations}
For each problem considered, the input flow rate is constrained such that $
    0 \leq u_k \leq 0.5 \,\; \forall k \in \{1,\ldots,N\}$,
and the optimization problems minimize the sum of the squares of the control inputs.
%\begin{equation}
%    J(x,u) = \sum_{k=1}^N u_k^2.
%\end{equation}

\textbf{Case 1:} Starting from $h^1_0 = 5$ and $h^2_0 = 5$, we find a control trajectory $u$ such that $h^1_N = 1.5$ and $h^2_N = 3.5$. Applying our binary-free reformulation to the constraints in \eqref{eq:h1_bar} and~\eqref{eq:h2_bar} allows the problem to be solved in Ipopt, yielding the trajectories shown in Figure~\ref{fig:two_tank_1}. Our solver converges on~74.1\% of runs exclusively to the optimal solution. We were unable to solve this problem using a Big M or complementarity-based differentiable reformulation. 

\textbf{Case 2:} Starting from $h^1_0 = 5$ and $h^2_0 = 2$, we find a control trajectory $u$ such that $h^1_N = 2$ and $h^2_N = 4$. We also require that the solution satisfies the logic constraint
\begin{equation}
    \label{eq:two_tank_case_2_logic}
    \left[ h_k^1 \geq 4.5 \right] \, \mathcal{U} \, \left[ h_k^2 \geq 4.5 \right] 
\end{equation}
which prevents the level in the first tank from dropping below 4.5 until the level in the second tank exceeds~4.5. We handle the constraints of \eqref{eq:h1_bar} and \eqref{eq:h2_bar} as in Case~1, and apply the method of Section \ref{sec:until_operator} to~\eqref{eq:two_tank_case_2_logic}. The optimal trajectory in Figure~\ref{fig:two_tank_2} is obtained for 38.3\% of runs, while 86.2\% of runs yield feasible solutions. Once again, the problem could not be solved using smooth big~M or complementarity-based reformulations.

\begin{table}[tb]
\centering
\caption{System Parameters for Two Tank Problem}
\label{tab:two_tank_param}
\begin{tabular}{ll|ll|ll}
\hline
$N$   & 20    & $T_s$    & 3 \,\text{s}   & $h_{s_{12}}$ & $2\,\text{m}$    \\ \hline
$A_1$ & $2 \,\text{m}^2$    & $A_2$    & $1 \,\text{m}^2$     & $h_{s_{2}}$  & $3 \,\text{m}$    \\ \hline
$s_1$ & $0.015 \,\text{m}^2$ \quad& $s_{12}$ & $0.02 \,\text{m}^2$ \quad& $s_2$        & $0.02 \,\text{m}^2$ \\ \hline
\end{tabular}
\end{table}

\section{Conclusions}

%In this paper, we presented a general framework for embedding discrete logic constraints into nonlinear optimization problems using exact smooth reformulations. By expressing arbitrary logic as max–min inequalities and applying a smoothing transformation to the result, our method enables complex logical requirements to be handled natively by continuous optimization solvers. Numerical results on quadrotor and two-tank problems illustrate the advantages of this approach including faster computation times and improved convergence. While the optimal choice of logical representation for smoothing remains an open question, and convergence rates suffer as the size of the logic constraints increases, the proposed method provides clear benefits for continuous optimization problems with moderate logical requirements.

In this paper, we presented a framework for embedding logic constraints into nonlinear optimization problems through exact binary-free reformulations. By expressing logic as max-min inequalities and smoothing the result, our method enables logical requirements to be handled directly by continuous optimization solvers. Numerical results on the quadrotor and two-tank problems demonstrate the advantages of the proposed reformulation, including faster computation and improved cost performance. Although the choice of logical representation remains an open question and convergence may degrade with large logic structures, the method offers clear benefits for continuous optimization problems with moderate logical complexity.

%\begin{ack}                               % Place acknowledgements
%This work was funded through a Natural Sciences and Engineering Research Council of Canada PGS D grant.   % here.
%\end{ack}

\bibliographystyle{plain}        % Include this if you use bibtex 
\bibliography{autosam}           % and a bib file to produce the 

@article{kirjner1998conversion,
  title={On the conversion of optimization problems with max-min constraints to standard optimization problems},
  author={Kirjner-Neto, C and Polak, Elijah},
  journal={SIAM Journal on Optimization},
  volume={8},
  number={4},
  pages={887--915},
  year={1998},
  publisher={SIAM}
}

@article{Lubin2023,
    author = {Miles Lubin and Oscar Dowson and Joaquim {Dias Garcia} and Joey Huchette and Beno{\^i}t Legat and Juan Pablo Vielma},
    title = {{JuMP} 1.0: {R}ecent improvements to a modeling language for mathematical optimization},
    journal = {Mathematical Programming Computation},
    year = {2023}
}

@article{wachter2006implementation,
  title={On the implementation of an interior-point filter line-search algorithm for large-scale nonlinear programming},
  author={W{\"a}chter, Andreas and Biegler, Lorenz T},
  journal={Mathematical Programming},
  volume={106},
  pages={25--57},
  year={2006},
  publisher={Springer}
}

@inproceedings{wehbeh2024robust,
  title={Robust Output Feedback of Nonlinear Systems through the Efficient Solution of Min-Max Optimization Problems},
  author={Wehbeh, Jad and Kerrigan, Eric C},
  booktitle={2024 IEEE 63rd Conference on Decision and Control (CDC)},
  pages={8870--8875},
  year={2024},
  organization={IEEE}
}

@book{boyd2004convex,
  title={Convex optimization},
  author={Boyd, Stephen P and Vandenberghe, Lieven},
  year={2004},
  publisher={Cambridge university press}
}

@article{wensing2023optimization,
  title={Optimization-based control for dynamic legged robots},
  author={Wensing, Patrick M and Posa, Michael and Hu, Yue and Escande, Adrien and Mansard, Nicolas and Del Prete, Andrea},
  journal={IEEE Transactions on Robotics},
  volume={40},
  pages={43--63},
  year={2023},
  publisher={IEEE}
}

@article{kirches2020approximation,
  title={Approximation properties and tight bounds for constrained mixed-integer optimal control},
  author={Kirches, Christian and Lenders, Felix and Manns, Paul},
  journal={SIAM Journal on Control and Optimization},
  volume={58},
  number={3},
  pages={1371--1402},
  year={2020},
  publisher={SIAM}
}

@article{kronqvist2019review,
  title={A review and comparison of solvers for convex {MINLP}},
  author={Kronqvist, Jan and Bernal, David E and Lundell, Andreas and Grossmann, Ignacio E},
  journal={Optimization and Engineering},
  volume={20},
  year={2019},
  publisher={Springer}
}

@article{liberti2019undecidability,
  title={Undecidability and hardness in mixed-integer nonlinear programming},
  author={Liberti, Leo},
  journal={RAIRO-Operations Research},
  volume={53},
  number={1},
  pages={81--109},
  year={2019},
  publisher={EDP Sciences}
}

@article{burer2012non,
  title={Non-convex mixed-integer nonlinear programming: A survey},
  author={Burer, Samuel and Letchford, Adam N},
  journal={Surveys in Operations Research and Management Science},
  volume={17},
  number={2},
  year={2012},
  publisher={Elsevier}
}

@article{malyuta2023fast,
  title={Fast homotopy for spacecraft rendezvous trajectory optimization with discrete logic},
  author={Malyuta, Danylo and A{\c{c}}{\i}kme{\c{s}}e, Beh{\c{c}}et},
  journal={Journal of Guidance, Control, and Dynamics},
  volume={46},
  number={7},
  pages={1262--1279},
  year={2023},
  publisher={American Institute of Aeronautics and Astronautics}
}

@article{bertsimas2021unified,
  title={A unified approach to mixed-integer optimization problems with logical constraints},
  author={Bertsimas, Dimitris and Cory-Wright, Ryan and Pauphilet, Jean},
  journal={SIAM Journal on Optimization},
  volume={31},
  number={3},
  pages={2340--2367},
  year={2021},
  publisher={SIAM}
}

@book{bazaraa2011linear,
  title={Linear programming and network flows},
  author={Bazaraa, Mokhtar S and Jarvis, John J and Sherali, Hanif D},
  year={2011},
  publisher={John Wiley \& Sons}
}

@article{belotti2016handling,
  title={On handling indicator constraints in mixed integer programming},
  author={Belotti, Pietro and Bonami, Pierre and Fischetti, Matteo and Lodi, Andrea and Monaci, Michele and Nogales-G{\'o}mez, Amaya and Salvagnin, Domenico},
  journal={Computational Optimization and Applications},
  volume={65},
  pages={545--566},
  year={2016},
  publisher={Springer}
}

@article{gunluk2010perspective,
  title={Perspective reformulations of mixed integer nonlinear programs with indicator variables},
  author={G{\"u}nl{\"u}k, Oktay and Linderoth, Jeff},
  journal={Mathematical Programming},
  volume={124},
  pages={183--205},
  year={2010},
  publisher={Springer}
}

@article{iliev2017approximation,
  title={On the approximation of the step function by some sigmoid functions},
  author={Iliev, A and Kyurkchiev, Nikolay and Markov, Svetoslav},
  journal={Mathematics and Computers in Simulation},
  volume={133},
  pages={223--234},
  year={2017},
  publisher={Elsevier}
}

@inproceedings{wolff2014optimization,
  title={Optimization-based control of nonlinear systems with linear temporal logic specifications},
  author={Wolff, Eric M and Topcu, Ufuk and Murray, Richard M},
  booktitle={Proc. of Int. Conf. on Robotics and Automation},
  pages={5319--5325},
  year={2014}
}

@article{belta2019formal,
  title={Formal methods for control synthesis: An optimization perspective},
  author={Belta, Calin and Sadraddini, Sadra},
  journal={Annual Review of Control, Robotics, and Autonomous Systems},
  volume={2},
  number={1},
  pages={115--140},
  year={2019},
  publisher={Annual Reviews}
}

@article{cafieri2023continuous,
  title={The continuous quadrant penalty formulation of logical constraints},
  author={Cafieri, Sonia and Conn, Andrew R and Mongeau, Marcel},
  journal={Open Journal of Mathematical Optimization},
  pages={1--12},
  year={2023}
}

@book{enderton2001mathematical,
  title={A mathematical introduction to logic},
  author={Enderton, Herbert B},
  year={2001},
  publisher={Elsevier}
}

@article{kurtz2022mixed,
  title={Mixed-integer programming for signal temporal logic with fewer binary variables},
  author={Kurtz, Vincent and Lin, Hai},
  journal={IEEE Control Systems Letters},
  volume={6},
  pages={2635--2640},
  year={2022},
  publisher={IEEE}
}

@inproceedings{papusha2016automata,
  title={Automata theory meets approximate dynamic programming: Optimal control with temporal logic constraints},
  author={Papusha, Ivan and Fu, Jie and Topcu, Ufuk and Murray, Richard M},
  booktitle={2016 IEEE 55th Conference on Decision and Control (CDC)},
  pages={434--440},
  year={2016},
  organization={IEEE}
}

@inproceedings{wehbeh2025smooth,
  author={J. Wehbeh and E. C. Kerrigan},
  title={Smooth Logic Constraints in Nonlinear Optimization and Optimal Control Problems},
  booktitle={Proceedings of the 64th IEEE Conference on Decision and Control (CDC)},
  year={2025},
}

@article{van1998constraint,
  title={Constraint programming over nonlinear constraints},
  author={Van Hentenryck, Pascal and Michel, Laurent and Benhamou, Fr{\'e}d{\'e}ric},
  journal={Science of computer programming},
  volume={30},
  number={1-2},
  pages={83--118},
  year={1998},
  publisher={Elsevier}
}
                                 % bibliography (preferred). The
                                 % correct style is generated by
                                 % Elsevier at the time of printing.

%\begin{thebibliography}{99}     % Otherwise use the  
                                 % thebibliography environment.
                                 % Insert the full references here.
                                 % See a recent issue of Automatica 
                                 % for the style.
%  \bibitem[Heritage, 1992]{Heritage:92}
%     (1992) {\it The American Heritage. 
%     Dictionary of the American Language.}
%     Houghton Mifflin Company.
%  \bibitem[Able, 1956]{Abl:56}
%     B.~C.~Able (1956). Nucleic acid content of macroscope. 
%     {\it Nature 2}, 7--9. 
%  \bibitem[Able {\em et al.}, 1954]{AbTaRu:54}   
%     B.~C. Able, R.~A. Tagg, and M.~Rush (1954).
%     Enzyme-catalyzed cellular transanimations.
%     In A.~F.~Round, editor, 
%     {\it Advances in Enzymology Vol. 2} (125--247). 
%     New York, Academic Press.
%  \bibitem[R.~Keohane, 1958]{Keo:58}
%     R.~Keohane (1958).
%     {\it Power and Interdependence: 
%     World Politics in Transition.}
%     Boston, Little, Brown \& Co.
%  \bibitem[Powers, 1985]{Pow:85}
%     T.~Powers (1985).
%     Is there a way out?
%     {\it Harpers, June 1985}, 35--47.

%\end{thebibliography}

\end{document}